\documentclass[epj]{svjour}
\usepackage{latexsym}
\usepackage{graphics}
\usepackage{amsmath}
\usepackage{array}
\usepackage{amssymb}
\usepackage{enumitem} 
\usepackage{lineno}
\usepackage{booktabs}
\usepackage{multicol}
\usepackage{multirow}
\usepackage{braket}
\usepackage{epstopdf}
\usepackage{hyperref}
\usepackage{xcolor}
\usepackage{cancel}
\usepackage[normalem]{ulem}
\usepackage[numbers,sort&compress]{natbib}
\newcommand{\acknowledgments}{\section*{Acknowledgments}}
\begin{document}
\title{Mass for Particles from Extra Dimensions }
\author{F. E. A. de Souza\inst{1}\thanks{\emph{francisco.emmanoel@uece.br}} \and M. O. Tahim\inst{1}\thanks{\emph{makarius.tahim@uece.br}} \and R. I. O. Júnior\inst{1}\thanks{\emph{raimundo.ivan@uece.br}} \and I. M. Macêdo\inst{1}
\thanks{\emph{metalisaak@gmail.com}}%
}                     
\offprints{}          
\institute{Universidade Estadual do Ceará, Faculdade de Educação, Ciências e Letras do Sertão Central\\
Quixadá-CE, Brazil.}
\date{Received: date / Revised version: date}
%
\abstract{We investigate a geometric mechanism for mass generation in a codimension-two hedgehog braneworld. A massless particle propagating in the bulk acquires an effective four-dimensional mass through its motion along the compact angular extra dimension, yielding a dynamically determined Kaluza–Klein spectrum after canonical quantization. While the bosonic sector exhibits no radial confinement, extending the analysis to N = 1, 2 spinning-particle models shows that the spin-curvature coupling modifies the effective potential and introduces a level-dependent localization pattern for the Kaluza–Klein modes. The results establish a particle-mechanics realization of geometric mass generation in flux-supported codimension-two braneworlds and highlight the interplay between bulk geometry, spin, and the effective four-dimensional mass spectrum.
\PACS{
      {PACS-key}{discribing text of that key}   \and
      {PACS-key}{discribing text of that key}
     } 
} 
\maketitle
\section{Introduction}
\label{intro}
The origin of mass remains one of the most profound questions in modern theoretical physics, bridging the gap between the mathematical elegance of gauge symmetries and the observed physical reality of massive particles. In the standard framework of quantum field theory, mass terms often appear to violate the underlying gauge or scale symmetries of the Lagrangian. Consequently, several mechanisms have been proposed to generate mass dynamically or geometrically without affecting the internal consistency of the theory.

The most empirically validated framework for mass generation is the Higgs Mechanism \cite{Higgs:1964pj}. In the Standard Model, the $SU(2)_L \times U(1)_Y$ electroweak symmetry prevents gauge bosons and fermions from possessing intrinsic mass. This is resolved by introducing a complex scalar field, the Higgs field, which acquires a non-zero vacuum expectation value (VEV) through a``Mexican hat'' potential. As the field settles into its minimum, the local gauge symmetry is spontaneously broken. The Goldstone bosons are ``eaten'' by the gauge fields, providing them with longitudinal degrees of freedom and, consequently, mass. For a gauge field $A_\mu$, the mass squared is proportional to the coupling $g$ and the VEV $v$: $M^2 = \frac{1}{4}g^2 v^2$.

Beyond spontaneous symmetry breaking, mass can em-
\newline
erge from the topological structure of the gauge fields themselves, particularly in lower-dimensional systems or effective field theories. In $(2+1)$ dimensions, the addition of a Chern-Simons term allows gauge bosons to become massive while maintaining explicit gauge invariance \cite{Deser:1982vy,Deser:1981wh}. Unlike the Higgs mechanism, this does not require a scalar sector. Similarly, in the BF model \cite{Allen:1990gb}, mass is generated through the topological coupling between a $(p)$-form gauge field and a $(d-p-1)$-form field. These mechanisms are central to our understanding of fractional quantum Hall states \cite{Tong:2016kpv} and topological insulators \cite{Hasan_2010}, where the ``mass'' manifests as an energy gap in the spectrum of excitations, rooted in the global properties of the manifold rather than local field dynamics.

A third paradigm suggests that mass is a consequence of the geometry of spacetime itself. In Kaluza-Klein (KK) theories, our four-dimensional world is viewed as a slice of a higher-dimensional manifold where the extra dimensions are compactified (e.g., on a circle $S^1$). The momentum of a field along these hidden dimensions appears to a four-dimensional observer as a discrete spectrum of masses \cite{Kaluza:1921tu,Klein:1926tv}:
$$
M_n^2 = \frac{n^2}{R^2}
$$
where $R$ is the compactification radius. Alternatively, the Randall-Sundrum (RS) models works with ``warped'' extra dimensions \cite{Randall:1999ee,Randall:1999vf}. By placing our 4D brane at a specific location within a 5D Anti-de Sitter (AdS) space, the gravitational hierarchy is addressed, and mass scales are shifted geometrically. In this context, the particle masses are determined by their overlap with the extra-dimensional geometry, offering a purely spatial solution to the hierarchy problem. The same behavior can be found to cases with dimensions $D>5$. There is particular interest in models where the branes are topological defects with several codimensions. In that cases, the membrane thickness appears naturally, making them more realistic.

All of these results are field theoretical ones. It would be interesting to build the analogue effects to particle action models. In particular, in this work, we discuss an analogue Kaluza-Klein compactification mechanism in the presence of codimension-2 membrane for pure particles. We show that a massless particle propagating in the bulk of codimension-2 hedgehog-type braneworld acquires an effective mass on the brane, generated purely by its motion along the extra angular dimension, $m=\gamma^{1/2}\dot\theta$. Mass here is a purely geometric consequence of the flux-supported bulk background.

The organization of this work is as follows. In the first section, we show that a particle in the $D=5$ usual Randall-Sundrum background does not present a kaluza-klein like mass interpretation which, together the behavior of non-confinement, breaks the idea in this dimensionality. Then, we review a codimension-2 hedgehog-type braneworld in preparation to study particles in this background. This is just what is made in the second section, where we find a kaluza-klein mass interpretation without, however, the confinement condition. Finally, in the third section, we show that the spinning particle models complete this proposal by ensuring confinement. We present, after all, conclusions and perspectives.


\section{Particles in Randall-Sundrum Backgrounds.}

In this section we will discuss the confinement of a bosonic test particle in a codimension one braneworld. In order to get a more general analysis, let us consider a generic background metric given by
\begin{eqnarray}\label{001}
ds^{2}=e^{2a(y)}\eta_{\mu\nu}dx^{\mu}dx^{\nu}+e^{2b(y)}dy^{2}.
\end{eqnarray}
The warp factors $a(y)$ and $b(y)$ are functions only of the extra coordinate $y$, which is considered infinitely large. In addition, $\eta_{\mu\nu}$ is the Minkowski metric with signature $(-,+,+,+)$. This generic shape of the metric allows to discuss a variety of braneworlds, among them, the delta-like RS-II model \cite{Randall:1999ee} and also some thick brane models \cite{Gremm:1999pj,Kehagias:2000au,Bazeia:2002xg,Bazeia:2004dh,Landim:2011ki}.

\subsection{The Bosonic Test Particle}

To start this discussion about the confinement of a bosonic test particle, let us use an alternative form for the action of a free relativistic particle. It is given by
\begin{eqnarray}\label{002}
S=\frac{1}{2}\int\left[e^{-2}g_{PQ}\dot{x}^{P}\dot{x}^{Q}-M^{2}\right]e d\tau.
\end{eqnarray}
In this action, $e$ is the vierbein, $\dot{x}^{P}=\frac{dx^{P}}{d\tau}$ and $M$ is the particle mass in $5$D. For the free case, the particle mass is a constant $M_{0}$. As we will show below, $e$ is not really a dynamic variable and it can be chosen arbitrarily. The vierbein choice means setting the parameterization (gauge condition) \cite{Souza:2019jqz}.

From the action (\ref{002}), we can obtain the equations of motion
\begin{eqnarray}
\frac{\delta S}{\delta e}=0 \to g_{PQ}\dot{x}^{P}\dot{x}^{Q}+e^{2}M^{2}(y)=0,\hspace{1.35cm} \label{005}\\
\frac{\delta S}{\delta x^{N}}=0 \to \frac{D}{D\tau}\left[e^{-1}\dot{x}^{N}\right]+\delta_{y}^{N}eMM'e^{-2b}=0.\label{006}
\end{eqnarray}

Choosing the gauge $e = e^{2a(y)}$ and using the Christoffel symbols of the metric above, the equations decouple into a conserved $4$-momentum
\begin{equation}
    p_\mu = e^{-2a} g_{\mu\nu} \dot{x}^\nu = \text{constant}
\end{equation}
and an equation for the extra-dimensional motion that can be written as an energy-conservation law,
\begin{equation}
    e^{(2b(y)-2a(y))} \dot{y}^2 + M(y)^2 e^{2a(y)} = C^2
\end{equation}
with $C^2 = p^2$. This allows the definition of an effective potential governing the motion in $y$:
\begin{equation}
U_{eff}(y) = M(y)^2 e^{2a(y)} - M(0)^2
\end{equation}

A particle is confined at $y = 0$ only if $U'_{eff}(0) = 0$ and $U''_{eff}(0) > 0$. The first condition requires $M'(0) = 0,$ and the second, using $a'(0) = 0$ (a standard feature of brane models), reduces to
\begin{equation}
    MM'' + M^2a''>0,
\end{equation}
remember that the above conditions are necessary but not sufficient to confine a particle \cite{Souza:2019jqz}. 

For a free particle $M(y) = M_0 = \text{constant}$ the condition above reduces to:
$$
a''(0)>0,
$$
this condition is never satisfied, since Einstein's equations for RS-like models with localized gravity always give $a''(0)$
\newline
$ < 0$. Therefore neither massive nor massless free bosonic particles can be trapped on the brane. This result reproduces, by a more general method valid also for $M = 0$, what was previously found for the delta-like RS-II model \cite{Mueck:2000bb,Dahia:2007ep,Souza:2019jqz}.
\subsubsection{Localization Mechanisms}

Since the free particle cannot be confined, an interaction is needed. A possible mechanism couples the particle to a scalar field $\phi(y)$ through
\begin{equation}
    M(y)^2 = M_0^2 + h^2\phi^2
\end{equation}
analogous to a Yukawa coupling \cite{Dahia:2007ep}. In this approach, the effective potential becomes
\begin{equation}\label{potda}
	U_{eff}(y)=\left[e^{2a}(M^{2}_{0}+h^{2}\phi^{2})-M^{2}_{0}\right],
\end{equation}
Applying the confinement conditions above to this mass function yields an upper bound on the confined mass,
\begin{equation}
    M_0 < \sqrt{3}\frac h\kappa
\end{equation}
where $\kappa$ is the $5D$ gravitational coupling constant. This bound can be made arbitrarily large by choosing $h$, but it still restricts which masses can be trapped.

An alternative mechanism couples the particle to the dilaton field $\pi(y)$, replacing the mass as \cite{Damour:1990tw,Kehagias:2000au,Alencar:2018cbk}, 
\begin{equation}
    M(y) = M_0 e^{\lambda\pi(y)},
\end{equation}
following the approach used in \cite{Damour:1990tw} where a particle with mass constant in a conformal Jordan frame is viewed in a Einstein frame to justify this coupling. The dilaton field is proportional to the warp factor, $\pi(y) = -\sqrt{3M_p^3}\,a(y)$ with $M_p$ the $5D$ Plank scale \cite{Kehagias:2000au}, the effective potential becomes.
\begin{equation}
    	U_{eff}(y)=M^{2}_{0}(e^{2a(1-\lambda\sqrt{3M^{(3)}_{p}})}-1)
\end{equation}
The confinement conditions then reduce to
\begin{equation}
    \lambda > \frac1{\sqrt{3M_p^3}}
\end{equation}
with no dependence on $M_0$. This mechanism therefore confines particles of any non-zero mass, unlike the scalar coupling, which only confines masses below a bound. Since $U_eff$ in this case is unbounded from above, this dilaton coupling provides a stronger localization mechanism.

\subsection{Spinning \texorpdfstring{$N=1,2$}{N=1,2} test particle.}

In order to include spin, Grassmann variables $\psi^P$ and $\xi$ are added following the Brink–Di Vecchia–Howe pseudomechanics formulation \cite{Brink:1976uf}
\begin{eqnarray}
		S=\frac{1}{2}\int\left[e^{-1}g_{AB}\dot{x}^{A}\dot{x}^{B}-eM^{2}_{0}-ig_{AB}\psi^{A}\frac{D\psi^{B}}{D\tau}\right.\nonumber\\
	\left.-i\xi\dot{\xi}-i\chi\left(e^{-1}g_{AB}\dot{x}^{A}\psi^{B}+M_{0}\xi\right)\right]d\tau,
\end{eqnarray}
where this action describes spinning particles $N=1$, and $\chi$ enforces local worldline supersymmtry. Fixing $e=e^{2a(y)}$ and $\chi = 0$ the extra dimension component of the modified geodesic equation, which couples spin to the Riemann tensor, leads to the effective potential \cite{Souza:2019jqz}
\begin{eqnarray}
	U_{eff}=M_{0}^{2}\left[e^{2a}-\frac{2i}{M_{0}}\xi\psi^{y}_{0} e^{a-b}a'\right].\label{424}
\end{eqnarray}
For the massless case $U_{eff} = 0$, so no confinement occurs, as in bosonic case. For the massive case, the spin term shifts the extremum of the potential away from $y = 0$. A general argument, independent of the warp factor, shows that $U'_{eff}$ at the brane position $y_B$ never vanishes, and that $U_{eff}(\infty) - U_{eff}(y_B) = -M_0^2 < 0$, so the potential at infinity is always lower at the brane. Consequently, any $N = 1$ spinning particle placed on the brane eventually escapes to the extra dimension \cite{Souza:2019jqz}.

The extension for $N = 2$ spinning particles with a non-confinement result was recently generalized to particles with extended worldline supersymmetry ($N = 2$): these are the classical objects whose quantization is associated with antisymmetric $p$-form fields \cite{deSouza:2025wzv}. Building an $N = 2$ pseudomechanics action with two independent sets of Grassmann variables \cite{Gershun:1979fb,deSouza:2025wzv}, the authors show that the same qualitative conclusion holds: the $N = 2$ spinning particle cannot be confined on the brane in the Randall-Sundrum scenario, for the same structural reason found in the $N = 1$ case, namely that the brane position is never a true minimum of the corresponding effective potential once the curvature coupling of the spin variables is included \cite{deSouza:2025wzv}. The authors further point out a ``satellite" behavior of the trajectory around the brane and argue that this non-confinement result has direct consequences for the localization of $p$-form fields, since these fields arise from the quantization of the $N = 2$ superparticle: if the classical superparticle itself is not trapped by the brane, this undermines standard expectations about localizing the corresponding $p$-form fields on the membrane.

\section{Bulk \texorpdfstring{$p$}{p}-forms.}
\label{confboson}

The presence of a bulk $p$-form field $A_{\mu_1...\mu_p}$ directly leads to a regular geometry in which stability may be insured simply by the magnetic flux conservation. The $D$-dimensional action is
\begin{eqnarray}
    S&=&\int d^Dx \sqrt{|g|}\Bigg[\frac12 M^{n+2}_D R-\frac{\Lambda_D}{M^{n+2}_D}+\nonumber\\
    &&(-1)^{p}\frac14 F_{\mu_1...\mu_{p+1}}F^{\mu_1...\mu_{p+1}}\Bigg].\label{pformact}
\end{eqnarray}
A solution for the equations of motion when $p=n-2$ is given by \cite{Gherghetta:2000jf}:
\begin{equation}
    F_{\theta_1 \dots\theta_{n-1}}=Q(\sin{\theta_{n-1}})^{n-2}\dots\sin{\theta_2}
\end{equation}
where $Q$ is the charge of the field configuration and the other $F$  components  are equal to zero. This
“hedgehog” field configuration is the generalization of the magnetic field of a monopole. The stress-energy tensor associated with this $p$-form field in the bulk is
\begin{equation}\label{stresspform}
T^{\mu}_{\nu}=\frac{(n-1)!}{4}\frac{Q^2}{\gamma^{n-1}}\delta^{\mu}_{\nu}\,;\,\, T^{r}_{r}=-T^{\theta}_{\theta}=\frac{(n-1)!}{4}\frac{Q^2}{\gamma^{n-1}}.
\end{equation}
The line element is given by \cite{Gherghetta:2000jf}
\begin{equation} \label{metricpform}   ds^{2}=\sigma(r)\eta_{\mu\nu}dx^{\mu}dx^{\nu}+dr^{2}+\gamma\,d\Omega^2_{n-1}.
\end{equation}
Then, the ansatz above, together with the stress-energy tensor \eqref{stresspform} , and the Einstein equations with $\Lambda_{\text{phys}}=0$ are reduced to the following equations for the factors outside the brane source:
\begin{eqnarray}
    &&(n-1)!\frac{Q^{2}}{\gamma^{n-1}}-\frac{1}{2\gamma}(n+2)(n-2)+\frac{\Lambda_D}{M_{D}^{n+2}}=0,\label{pformgamma}\\
    &&c^{2}=-\frac12 \frac{\Lambda_D}{M_{D}^{n+2}}+\frac{1}{4\gamma}(n-2)^{2},\label{pformc2}
\end{eqnarray}
where $c^{2}>0$ and $\gamma>0$ do not change the signature of the metric, with
\begin{equation}
    c=\sqrt{\frac{-\Lambda_6}{2M_{6}^{4}}}\label{constc},
\end{equation}
for codimension $n=2$. The equations \eqref{pformgamma} and \eqref{pformc2} show that, for $n=2$, gravity is localized only if the bulk cosmological constant $\Lambda_6$ is negative and the action  \eqref{pformact} describes a scalar field ($0$-form). Where it is assumed a solution outside the core of the form
\begin{equation}
    \sigma(r)=e^{-cr} \,\,\text{and}\,\, \gamma = \frac{Q^2}{2 c^2},\label{sign}
\end{equation}
with $\gamma$ constant. For $n=2$ the background field is a $0$-form with winding flux; it only supports the geometry \cite{Gherghetta:2000jf}. The fields associated with the particle models are test fields: scalar (spinless), spinor ($N=1$) and the field strength of a ($q-1$)-form potential ($N=2$) \cite{Brink:1976sc}. Then, for a particle in this scenario near the brane region  and with $n = 2$, the metric \eqref{metricpform} take the form
\begin{equation}\label{metricform2}    ds^{2}=\sigma(r)\eta_{\mu\nu}dx^{\mu}dx^{\nu}+dr^{2}+\gamma\,d\theta^2,
\end{equation}
that give us the following Christoffel symbols:
\begin{equation}
 \Gamma^{\mu}_{r\nu}=\frac{1}{2}\frac{\sigma'(r)}{\sigma(r)}\delta^{\mu}_{\nu}\,,\,\,\Gamma^{r}_{\mu\nu}=-\frac{1}{2}\sigma'(r)\eta_{\mu\nu}.\label{gammapform2}
\end{equation}

\subsection{Codimension \texorpdfstring{$n=2$}{n=2}: Spinless particles}

 Let us now, in  this background,  analyze the behavior of a massless particle. To this, we take the action for this massless particle to be Polyakov like, i.e 
\begin{equation}\label{actionE}
    S=\frac{1}{2}\int e^{-1}g_{AB}\dot{x}^{A}\dot{x}^{B}d\tau.
\end{equation}
With the metric tensor $g_{AB} = \text{diag}[\sigma(r)\eta_{\mu\nu},1,\gamma]$,
we obtain the equations of motion
\begin{eqnarray}
    &&\frac{\delta S}{\delta e}\rightarrow g_{AB}\dot{x}^{A}\dot{x}^{B}=0\label{motE}\\
    && \frac{\delta S}{\delta \dot{x}^{C}}\rightarrow \frac{D}{D\tau}[e^{-1}g_{AC}\dot{x}^{A}]=0.\label{motX}
\end{eqnarray}
Where the equation for $e$ is not a dynamical one. Then, using the Christoffel symbols \eqref{gammapform2} and fixing the gauge as $e=1$, we obtain the following  equations  for $x^{A}$ 
\begin{eqnarray}
&&\frac{D}{D\tau}[\dot{x}^{\mu}]=0\rightarrow\ddot{x}^{\mu}+                                      \frac{\sigma(r)'}{\sigma(r)}\dot{x}^{\mu}\dot{r}=0 \label{mot1pform} \\
&&\frac{D}{D\tau}[\dot{r}]=0\rightarrow\ddot{r}-
    \frac{1}{2}\frac{\sigma(r)'}{\sigma(r)}g_{\mu\nu}\dot{x}^{\mu}\dot{x}^{\nu}=0\label{mot2pform}\\
&&\frac{D}{D\tau}[\dot{\theta}]\equiv
    \frac{d}{d\tau}\dot{\theta}=0\rightarrow\ddot{\theta}=0,\label{mot3pform}   
\end{eqnarray}
and the constraint
\begin{equation}    \sigma(r)\eta_{\mu\nu}\dot{x}^{\mu}\dot{x}^{\nu}+\dot{r}^2+\gamma\dot{\theta}^{2}=0.\label{constpform2}
\end{equation}
If we multiply \eqref{mot1pform} by $\sigma(r)$ and \eqref{mot2pform} by $\dot{r}\sigma(r)$ together with the constraint \eqref{constpform2}, we obtain conserved quantities in the $x^{\mu}$, $r$ and $\theta$ directions given, respectively, by:
\begin{eqnarray}
&& \frac{d}{d\tau}[\sigma(r)\dot{x}^{\mu}]=0\rightarrow \sigma(r)\dot{x}^{\mu}\equiv P^{\mu}(r),\\
&& \frac{d}{d\tau}\Bigg[\frac12\sigma(r)(\dot{r}^{2}+\gamma\dot{\theta}^{2})\Bigg]=0\nonumber\\
&&\qquad\rightarrow\frac12\sigma(r)(\dot{r}^{2}+\gamma\dot{\theta}^{2})\equiv C(r),\\
&& \frac{d}{d\tau}\dot{\theta}=0\rightarrow \dot{\theta} \equiv w.\label{thetamom}
\end{eqnarray}
The quantity $P^{\mu}(r)$ is constant on the hypersurface for constant $r$, so that for $r=0$ we have the conserved four-momentum $p^\mu=P^\mu(r)|_{r=0}$ on the brane. 

Using the constraint \eqref{constpform2}, the conserved momentum $P^\mu = \sigma(r)\dot x^\mu$ satisfies
\begin{equation}
    \eta_{\mu\nu}P^\mu P^\nu = - \sigma(\tilde{r}) [\dot{\tilde{r}}^2 + \gamma\dot{\theta}^2] = - 2C(\tilde{r}).\label{masshs}
\end{equation}
which is independent of $r$. At $r=0$, where $p^\mu = P^\mu(r) |_{r=0}$,
\begin{equation}
    \eta_{\mu\nu}p^\mu p^\nu = - (\dot{{r_0}}^2 + \gamma\dot{\theta}^2) \equiv - M^2,\label{massbr}
\end{equation}
with $m = \gamma^{1/2}\dot\theta$ the contribution of the compact direction, which is the only one quantized (Sec. \ref{kkspec:spin0}). For vanishing radial momentum, $M = m$. The two contributions to the invariant $M^2$ differ in nature. The angular term, $m^2 = \gamma\dot\theta ^2$, is quantized exactly and topologically: since $\theta$ is compact, single-valuedness of the wave function fixes $p_\theta = \gamma\dot\theta ^2 = l \in \mathbb{Z}$, which gives the discrete tower of Sec.\,\ref{kkspec:spin0}. For the radial term, $\dot r_0^2$, there is no analogous topological condition, since $r$ is non-compact. Its spectrum would follow from the radial wave equation obtained by quantizing the mass-shell constraint, and depends on the effective potential and on the boundary conditions. We expect it to be continuous when the radial motion is unbounded, as in the spinless case discussed below, and to be possibly discrete only if the motion is bound, as can happen for spinning particles (Sec. \ref{confspin}). In that case the total spectrum would take the form $M^2 = M^2_{rad} + l^2/\gamma$, which reduces to the tower $m_l = l\gamma^{-\frac12}$ for modes without radial momentum. The radial dynamics, governed by the effective potential \eqref{ueffcd2} and, with spin, by \eqref{effn2}, controls the localization of the particle rather than the discrete KK spectrum. The explicit quantization of the radial sector is left for future work. 

For the extra radial dimension, $C(r)$ is a constant along it. Then, evaluating $C(r)|_{r=0}=C(r)$ and organizing the terms we find
\begin{equation}
    \sigma(r)\dot{r}^2 = \dot{r}^2_0-m^2[\sigma(r)-1].
\end{equation}
This last equation tells us that the particle has $\dot{ r }_{0}^{2}$ as the amount of initial energy along of the extra radial dimension. Then the effective potential acts on the particle such that since $\sigma(r)\dot r^2 \geq 0$, the particle can only move in the region where $\dot r_0^2 - m^2 u_{eff}(r) \geq 0$, that is, $U_{eff} \leq \dot r_0^2$, as in \cite{Dahia:2007ep}.
\begin{equation}\label{ueffcd2}
  U_{eff}( r )=m^{2}[\sigma( r )-1],
\end{equation}
where we define an effective potential density by $u_{eff}( r )={U_{eff}( r )}/{m^2}$ as:
\begin{equation}\label{ueffpform}
  u_{eff}(r)=\sigma(r)-1.
\end{equation}
Expanding the potential near the origin gives
\begin{equation}
u_{eff} \approx -cr + \frac{c^2}{2}r^2 + \dots,
\end{equation}
that is analogous to the effective potential density obtained in \cite{deSouza:2026pja}. So, we follow the same steps done  in it.
The particle is subject to the effective force density $f_{eff} = -u_{eff}'$. The force on the particle at the origin  is
\begin{equation}
f_{eff}(0) = c > 0,
\end{equation}
and then it is repelled from there. For arbitrary values of $r$ the effective force is given by
\begin{equation}
f_{eff} = c e^{-cr} > 0, \quad \forall  r > 0.
\end{equation}
Hence, the effective potential is strictly decreasing from $u_{eff}(0) = 0$ to
\begin{equation}
u_\infty = \lim_{r \rightarrow \infty} u_{eff}(r) = -1.
\end{equation}
\begin{figure}
\resizebox{0.5\textwidth}{!}{%
  \includegraphics{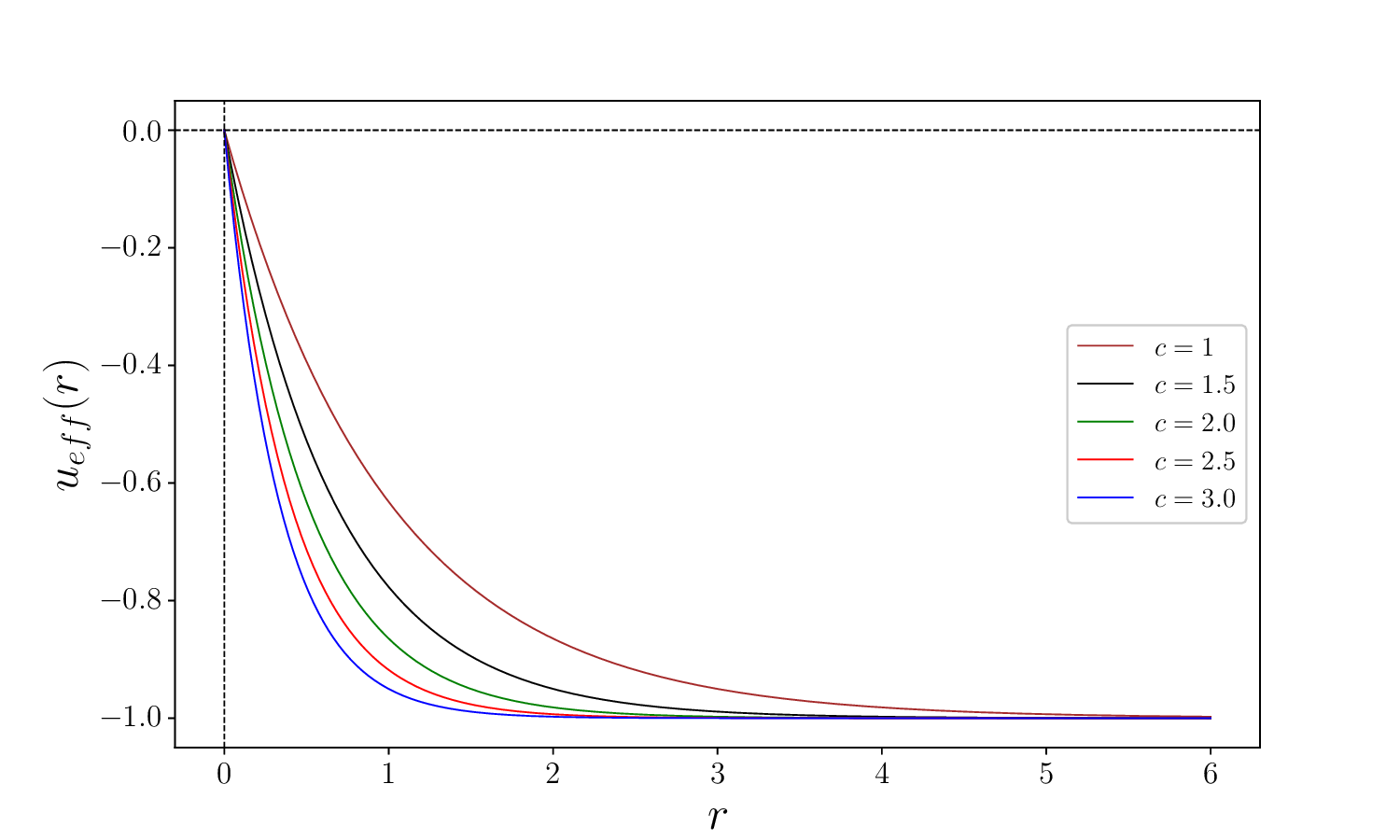}
}
\caption{The effective Hedgehog potential for various values of $c$}
\label{fig:spin}       
\end{figure}
Therefore, there are no critical points for $r > 0$, and the origin is not an equilibrium point since the force is nonzero. Consequently, there is no stable equilibrium anywhere in the system. For a classical particle, the force is always positive and points outward. A particle placed at any finite radius will accelerate towards larger $r$. If the particle comes from infinity with total energy $E > -1$, it will be reflected at the point where $u_{eff} = E$ and return to infinity — a scattering trajectory. The conclusion is that confinement does not occur.

\subsection{Connection to Kaluza-Klein spectra}
\label{kkspec:spin0}

So far $ \dot{\theta} $ has been treated as an arbitrary real constant of motion, leading to a continuous mass spectrum $ m=\gamma^\frac12 \dot{\theta}$. This single-valuedness condition can be established unambiguously by first
considering the flat-space limit of the metric \eqref{metricform2}, $\sigma(r)\to1$,
$\gamma\to1$, describing a free particle on $\mathbb{R}^{1,3}\times S^1$. In this
limit the action \eqref{actionE} reduces to that of a free particle on a flat cylinder, for
which $p_\theta=\dot\theta$ is the canonical momentum conjugate to $\theta$, and
canonical quantization together with the periodicity of the compact coordinate,
$\theta\sim\theta+2\pi$, forces the standard result $p_\theta=l\in\mathbb{Z}$. Since
this condition is topological -- it follows from the range of $\theta$ alone, not
from the dynamics -- it is preserved under minimal coupling to the curved
background, $\eta_{AB}\to g_{AB}=\mathrm{diag}(\sigma(r)\eta_{\mu\nu},1,\gamma)$,
which affects only the relation between $\dot\theta$ and the canonical momentum,
without altering the identification $\theta\sim\theta+2\pi$ itself. In the curved
background considered here, since the transverse metric for $n=2$ in \eqref{metricform2} is
$\gamma\, d\theta^2$, the canonical momentum conjugate to $\theta$ becomes
\begin{equation}
    p_\theta = \gamma \dot\theta,
\end{equation}
and is subject to the single-valuedness condition on the associated wavefunction,
$$
\Psi(\theta + 2\pi) = \Psi(\theta),
$$
upon canonical quantization of the action \eqref{actionE}. This forces
\begin{equation}
    p_\theta = l,\,\,\forall \,\,l\in\mathbb{Z}.
\end{equation}
and consequently the brane mass spectrum becomes discrete,
\begin{equation}
    m_l = \gamma^{-\frac12}\,l, \,\text{for}\,\, l=0,\pm1,\pm2,\dots\,.\label{discmass}
\end{equation}
This is precisely the Kaluza-Klein spectrum $ M_n = \frac nR $ quoted in the Introduction, with an effective compactification radius
\begin{equation}
    R_{eff} = \gamma^\frac12 = \frac{Q}{\sqrt2 \,c}.
\end{equation}
Unlike standard Kaluza-Klein compactifications, where $R$ is an external geometric input, here $R_{eff}$ is dinamically fixed by the flux charge $Q$ and by the bulk cosmological constant through $c$ \eqref{constc} — the compactification scale of the resulting tower is determined by the Einstein–flux solution itself.  Furthermore, since the warp factor $\sigma(\tilde r)$ multiplies the angular contribution to the conserved invariant, the angular part of each level is rescaled according to the position $\tilde r$ at which it is evaluated (the total invariant $M^2=\dot r_0^2+m^2$ being independent
of $\tilde r$),
\begin{equation}
    M_l(\tilde{r}) = \Bigg({\frac{\sigma(\tilde{r}) }{\gamma}}\Bigg)^\frac12\,l,
\end{equation}
a feature with no direct analog in either
pure Kaluza-Klein or pure Randall-Sundrum constructions.

\section{Bulk p-forms: Spinning particles.}
\label{confspin}
In this section, we will analyze the massless particle action for a spinning particle $N=1, 2$ \cite{Brink:1976uf, Brink:1976sc, Gershun:1979fb, Howe:1989vn, Howe:1988ft, Siegel:1988ru, Rivelles:1990dq} for a codimension $n=2$ topological defect background.  We start by taking the following action
\begin{eqnarray}\label{fullspinac}
    S&=&\int\Bigg[\frac{e^{-1}}{2}g_{AB}\dot{x}^{A}\dot{x}^{B}-\frac{i}{2}g_{AB}\psi^{A}_{k}\frac{D\psi^{B}_{k}}{D\tau}\nonumber\\
    &-&\frac{i}{2}e^{-1}\lambda_{k}g_{AB}\psi^{A}_{k}\dot{x}^{B}\Bigg.-\Bigg.\frac{1}{8}e^{-1}g_{AB}(\lambda_{k}\psi^{A}_{k})(\lambda_{k}\psi^{B}_{k})\nonumber\\
    &+&f\left[-i\frac{1}{2}\epsilon_{kl}g_{AB}\psi^{A}_{k}\psi^{B}_{l}-\Big(q-\frac{1}{2}D\Big)\right]\Bigg]d\tau,
\end{eqnarray}
for massless spinning particles in the bulk, where $k,l=1,2$ and we have a sum under which case. $N=1$ must represent a description of the spin particle $\frac12$ and $f=0$ because $k=1$, and we do not need the Chern-Simon term \cite{Souza:2019jqz, Brink:1976uf}. 

The case $N=2$ must represent a description of the gauge field \cite{deSouza:2025wzv, Howe:1989vn}. Without the Chern-Simons term, the $N = 2$ wavefunction is necessarily a $\frac12 D$-form, which is further constrained to be harmonic. This means that for $D$ odd the wavefunction must vanish. With the inclusion of $( q-\frac 12D)$ the wavefunction becomes a harmonic $q$-form. Since $q$ is an integer this means that for $D$ odd a non-vanishing wavefunction requires a non-vanishing Chern-Simons coefficient \cite{Howe:1989vn}. For $N=2$, the constraints imposed by the worldline supergravity fields are equivalent to
$\partial_{[\mu_1}F_{\mu_2\ldots\mu_{q+1}]}=0$ and $\partial^{\mu_1}F_{\mu_1\ldots\mu_q}=0$ \cite{Howe:1989vn}. The first is solved by $F=q\,\partial_{[\mu_1}A_{\mu_2\ldots\mu_q]}$,
which is the origin of the gauge potential $A$ and of its invariance $A\to A+d\Lambda$. The Chern--Simons-type term fixes the rank $q$ of the form and enters only through quantization.

It is well known that, in this case, the Grassmann variables contribute to the effective potential \cite{Souza:2019jqz, deSouza:2025wzv}. The question we pose is whether their contributions are relevant to the problem of particle confinement in this situation.

By the Hamilton principle, we obtain from (\ref{fullspinac}) the following equations of motion:
\begin{eqnarray}
\frac{\delta S}{\delta e} &\to& g_{AB}\left[\dot{x}^{A}\dot{x}^{B}-i\lambda_{k}\dot{x}^{A}\psi^{B}\right.\nonumber\\
&&\left.-\frac{1}{4}(\lambda_{k}\psi^{A}_{k})(\lambda_{k}\psi^{B}_{k})\right] = 0,\label{402}\\
\frac{\delta S}{\delta \lambda_{k}} &\to & g_{AB}\left[\dot{x}^{A}\psi^{B}-\frac{i}{2}(\lambda_{k}\psi^{A}_{k})\psi^{B}_{k}\right]=0,\label{404}\\
\frac{\delta S}{\delta f} &\to& \frac{1}{2}\epsilon_{kl}g_{AB}\psi^{A}_{k}\psi^{B}_{l}-i\Big(q-\frac{1}{2}D\Big)=0\label{409}\\
\frac{\delta S}{\delta \psi^{B}_{k}} &\to& \frac{D\psi^{B}_{k}}{D\tau}-\frac{1}{2}e^{-1}\lambda_{k}\dot{x}^{B}+\frac{i}{4}e^{-1}\lambda_{k}(\lambda_{k}\psi^{B}_{k})\nonumber\\
&+&f\epsilon_{lk}\psi^{B}_{l}=0,\label{403}\\
\frac{\delta S}{\delta x^{A}} &\to& \frac{D}{D\tau}\left[e^{-1}g_{AF}\dot{x}^{F}\right]-\frac{i}{2}\frac{d}{d\tau}\left[\lambda_{k} e^{-1}g_{AF}\psi^{F}_{k}\right]\nonumber\\& &+\frac{i}{2}R_{ASQR}\psi^{Q}_{k}\psi^{R}_{k}\dot{x}^{S}=0.
\label{406}
\end{eqnarray}

If we choose the gauge conditions on equations (\ref{402})-(\ref{406}) as $e=f=1$ and $\lambda_{k} = 0 $, we simplify the equations of motion as follows:
\begin{eqnarray}
&&\frac{D\psi^{P}_{k}}{D\tau}=0,\label{503}\\
&&\frac{D\dot{x}^{N}}{D\tau}+\frac i2 R^{N}_{\ SQR}\psi^{Q}\psi^{R}\dot{x}^{S}=0.\label{506}
\end{eqnarray}
The constraints are given by
\begin{eqnarray}
&&g_{PQ}\dot{x}^{P}\dot{x}^{Q}=0,\label{con1}\\
&&g_{PQ}\dot{x}^{P}\psi^{Q}_{k}=0,\label{con2}\\
&&g_{AB}\psi^{A}_{k}\psi^{B}_{l}=0\label{con3},
\end{eqnarray}

in order to study the generalization of the magnetic field of
a monopole for codimension $n=2$. Then, we can analyze the behavior of the spinning particle $N=1, 2$ in this background. In this sense, we need to obtain an effective potential to describe, as in \cite{deSouza:2025wzv}, the behavior of the particle in the region near the membrane.

\subsection{Codimension \texorpdfstring{$n=2$}{n=2}: Spinning Particle}\label{spincd2}

We will adopt the metric \eqref{metricform2} to describe the background of a massless spinning particle $ N = 1,2 $ in the bulk. Thus, the Christoffel symbols are given by \eqref{gammapform2}. 
Now, to obtain the motion equations for the $x^P$ direction in \eqref{506}, we need to obtain the Riemann tensor components, which are given by:
\begin{eqnarray}
    &&R^{\mu}_{\nu\kappa\tau}=\frac{1}{4}\frac{\sigma'^{2}( r )}{\sigma^2(r)}\Big[\delta^{\mu}_{\tau}g_{\nu\kappa}-\delta^{\mu}_{\kappa}g_{\nu\tau}\Big];\label{ric1}\\
    &&R^{\mu}_{rr\nu}=\left[\frac{1}{2}\frac{\sigma''( r )}{\sigma( r )}-\frac{1}{4}\frac{\sigma'^{2}( r )}{\sigma^{2}( r )}\right]\delta^{\mu}_{\nu};\label{ric2}\\
    &&R^{ r }_{\mu\nu  r }=\left[\frac{1}{2}\frac{\sigma''( r )}{\sigma( r )}-\frac{1}{4}\frac{\sigma'^{2}( r )}{\sigma^2( r )}\right]g_{\mu\nu};\label{ric4}\\
\end{eqnarray}
Now, taking equation \eqref{503} for $\mu$, $r$, and $\theta$, respectively, we obtain 
\begin{eqnarray}
    &&\dot{\psi\,^\mu _k} + \frac12 \frac{\sigma'}{\sigma}( \psi_k ^ r  \dot{x}^\mu + \psi_k ^\mu \dot{ r }) = 0 \label{psiu};\\
     &&\dot{\psi^ r  _k} + \frac12 \frac{\sigma'}{\sigma}(\psi_k ^ r  \dot{ r } + \gamma\dot{\theta}\psi_k ^\theta) = 0 \label{psip};\\
     &&\dot{\psi}^\theta = 0,\label{psiteta}
\end{eqnarray}
where we use the constraint \eqref{con2} in \eqref{psip}. Multiplying \eqref{psiu} and \eqref{psip} by $\psi^ r  _k$ and $\psi^\mu _k$, respectively, we find the result
\begin{equation}
   \frac{d}{d\tau}(\sigma \psi^ r _k \psi^\mu_k) + \frac12 \sigma'\xi_k\,\psi^\mu_k = 0, \label{psipu}
\end{equation}
where 
\begin{equation}
    \xi_k = \gamma\dot{\theta}\psi_k^\theta\label{xik}
\end{equation}
is a Grassmannian constant. Multiplying \eqref{psip} by $ \xi_k $ and adding them up, we obtain the  relation
\begin{equation}
    \psi^r_k\,\xi_k = \sigma^{-\frac12}\psi^r_k(0)\,\xi_k. \label{contxir}
\end{equation}
Thus, using \eqref{gammapform2}, \eqref{constpform2}, \eqref{psipu}, \eqref{contxir}, \eqref{ric1}-\eqref{ric4} in \eqref{506}, we get the conserved quantities for $ \mu $, $ \theta $ and $r$ given by:
\begin{eqnarray}
 &&\frac{d}{d\tau}(\sigma \dot{x}^\mu + \frac i2 \psi^r_k \psi^\mu_k\sigma') = 0\nonumber\\
 &&\qquad\rightarrow \sigma \dot{x}^\mu + \frac i2 \psi^r_k \psi^\mu_k\sigma' \equiv P^\mu (r),\label{consx} \\
 && \frac{d}{d\tau}[\sigma(\dot{r}^2 + \gamma\dot{\theta}^2)-i\xi_k \psi^r_k(0)\sigma'\sigma^{-\frac12}] = 0 \nonumber\\
 &&\qquad \rightarrow \sigma(\dot{r}^2 + m^2)-i\xi_k \psi^r_k(0)\sigma'\sigma^{-\frac12} \equiv Q(r), \label{consr}\\
 &&\frac{d}{d\tau}\dot{\theta} = 0 \rightarrow \dot{\theta} \equiv w\label{consteta},
\end{eqnarray}
where we define the conserved momentum
\begin{equation}
    P^\mu(r) = \sigma(r) \dot{x}^\mu + \frac i2 \psi^r_k \psi^\mu_k\sigma'(r).\label{consmomentum}
\end{equation}

As in the previous section, for the extra angular dimension, we have a mass creation starting from the motion of this massless particle along the extra angular dimension, once $\dot{\theta}$ is constant along the extra angular dimension, as shown in \eqref{thetamom}, with the angular velocity of the particle along the extra angular dimension  being a constant.

By equation \eqref{consr}, we have the conserved quantity
\begin{equation}
    Q(r) = \sigma(\dot{r}^2 + m^2)-i\xi_k \psi^r_k(0)\sigma'\sigma^{-\frac12},\label{Qr}
\end{equation}
which is constant. Then, if we compare $Q(r) = Q(r)|_{r=0}$ we get
\begin{eqnarray}
   &&\sigma(\dot{r}^2 + m^2)-i\xi_k \psi^r_k(0)\sigma'\sigma^{-\frac12} = \dot{r_0}^2 + m^2-i\xi_k \psi^r_k(0)\sigma'(0)\nonumber\\
   &&\sigma \dot{r}^2 = \dot{r_0}^2 \nonumber\\
   &&\qquad- m^2\Bigg[ \sigma - \frac i{m^2} \xi_k \psi^r_k(0)\Big(\sigma'\sigma^{-\frac12}-\sigma'(0)\Big) -1 \Bigg]\label{pot},
\end{eqnarray}
this last equation tells us that the particle has $\dot{ r }_{0}^{2}$ as the amount of initial energy along the extra radial dimension. Then, the last term in \eqref{pot} works as an effective potential acts on the particle such that since $\sigma(r)\dot r^2 \geq 0$, the particle can only move in the region where $\dot r_0^2 - m^2 u_{eff}(r) \geq 0$, that is, $U_{eff} \leq \dot r_0^2$, as in \cite{Dahia:2007ep} the effective potential density is given by
\begin{equation}\label{effn2}
u_{eff} =  \sigma - A\Big[\sigma'\sigma^{-\frac12}-\sigma'(0)\Big] -1,
\end{equation}
where 
\begin{equation}
    A=\frac i{m^2} \xi_k \psi^r_k(0).\label{spinprm}
\end{equation}
The parameter $A$ is a bilinear in Grassmann variables and is
therefore nilpotent at the classical level, so conditions such as
$A<0$ cannot be read as inequalities between genuine real numbers.
Following Refs.\cite{Souza:2019jqz,deSouza:2025wzv,deSouza:2026pja}, we treat $A$ as
an effective real parameter, i.e.\ the value of the bilinear in a given
spin state. Upon quantization the bilinear becomes an operator built
from the spin tensor, $S^{\theta r}\sim\psi^\theta\psi^r$, and the
conditions below become conditions on its eigenvalues, selecting the
relative orientation of the angular and radial spin components.
Confinement of the spinning particle is to be understood in this
sense, as in \cite{Souza:2019jqz,deSouza:2025wzv,deSouza:2026pja}: it occurs
    for spin configurations with $A<0$. When $\psi^N_k$ is zero, $A=0$, we recover the results discussed previously for a bosonic test particle given by equations \eqref{ueffpform}. 

We obtain, due to the spin variable, a new piece in $u_{eff}$ acting on the particle: the spin modifies the effective potential through its interaction with the curvature. This opens up the possibility of confining a spinning particle \cite{Souza:2019jqz, deSouza:2025wzv, deSouza:2026pja}.
We proceed to check that now, where the effective potential density  analyzed is analog to that obtained in \cite{deSouza:2026pja}. For $\sigma(r)$ given by \eqref{sign} and near the origin, the potential expands as 
\begin{equation}
    u_{eff}(r) \approx \Bigg(-c - \frac{Ac^2}{2} \Bigg)r + \Bigg(\frac{c^2}{2} + \frac{Ac^3}8\Bigg)r^2,
\end{equation}
where the constant $c$ is given by \eqref{constc}. The force density on the particle is $f_{eff}(r) = -u'_{eff}(r) $:
\begin{equation}
    f_{eff} = \Bigg(c + \frac{Ac^2}{2} \Bigg) - \Bigg(c^2 + \frac{Ac^3}4\Bigg)r.\label{densforce}
\end{equation}
At $r=0$
\begin{equation}
    f_{eff}(0) = c + \frac{Ac^2}{2}.
\end{equation}
The sign of $f_{eff}(0)$ determines the behavior of the particle near the origin:
\begin{enumerate}[label=(A.\Roman*), ref=A.\Roman*]
    \item \label{item:repulsive} If $f_{eff}(0)>0 \rightarrow A>-\frac{2}{c}$, the force points outward, and the particle is repelled from the origin;
    \item \label{item:attractive} If $f_{eff}(0)<0 \rightarrow A<-\frac{2}{c}$, the force points inward, and the particle is attracted toward the origin;
    \item \label{item:harmonic} If $f_{eff}(0)=0 \rightarrow A=-\frac{2}{c}$, the linear term vanishes, and the potential becomes harmonic.
\end{enumerate}
We should stress that, when the force at the origin vanishes, the effective potential near $r=0$ becomes
\begin{equation}
    u_{eff}(r) \approx \frac{c^2}{4}r^2.  
\end{equation}
This is the potential of the harmonic oscillator. The origin is a point of stable equilibrium: if the particle is slightly displaced, a restoring force $f_{eff}(r) = -\frac{c^2}2 r$ pulls it back, causing it to oscillate around $r=0$. Locally, the particle remains attained near the center. For large distances, the potential does not diverge. Instead, it approaches a finite constant
\begin{equation}
u_\infty = \lim_{r\rightarrow\infty} u_{eff}(r) = -1-Ac.  \label{uinfty}
\end{equation}
We now evaluate the critical points in this case. The derivative of the effective potential \eqref{effn2} is:
\begin{equation}
    u'_{eff} = -ce^{-\frac{cr}{2}}\Bigg(e^{-\frac{cr}{2}}+\frac{Ac}{2}\Bigg).
\end{equation}
Critical points occur when $u'_{eff} = 0$. Since $e^{-\frac{cr}{2}}$ for finite $r=\tilde{r}$, we require
\begin{equation}
    e^{-\frac{c\tilde{r}}{2}}+\frac{Ac}{2} = 0 \implies e^{-\frac{c\tilde{r}}{2}} = -\frac{Ac}{2},\label{minu}
\end{equation}
then, the effective potential has extrema if $A < 0$. For $A \geq 0$, there are no extremal points for this potential. Then, it has extremal points when
\begin{equation}
    \tilde{r} = -\frac2c \ln\Bigg(-\frac{Ac}{2}\Bigg),\, \text{for}\, A<0. \label{rtilde}
\end{equation}
where $\tilde{r}$ is obtained by \eqref{minu}.
To evaluate whether points are minimum, we need to check the second derivative of the effective potential
\begin{equation}
    u''_{eff} = c^2e^{-\frac12 cr}\Bigg(e^{-\frac12 cr} + \frac14 Ac \Bigg).
\end{equation}
Then, for $A=-\frac2c$, the origin $r=0$ is a minimum. The result in \eqref{rtilde} gives also a minimum point, since $-\frac2c < A < 0$, as
\begin{equation}
    r_{min} = -\frac2c \ln\Bigg(-\frac{Ac}{2}\Bigg)> 0,\, \text{for}\, -\frac2c<A<0. \label{rmin}
\end{equation}
This regime is characterized by a change in the sign of the force. Near the origin, the force is repulsive for $A>-\frac2c$. For large distances, the force becomes attractive when $A < 0$. The transition from repulsion to attraction gives rise to a local minimum at $r_{min} > 0$, corresponding to a stable equilibrium point. Then the conditions for equilibrium and stability of the spinning particle can be resumed as follows:
\begin{enumerate}[label=(B.\Roman*), ref=B.\Roman*]
    \item \label{item:stable-origin-strict} For $A < -\frac2c$, the origin is a region of stable equilibrium: small displacements result in a restoring force;
    \item \label{item:stable-origin-harmonic} For $A = -\frac2c$, the origin is a region of stable equilibrium (harmonic oscillator behavior);
    \item \label{item:stable-rmin} For $-\frac2c < A < 0$, there is no equilibrium at the origin, but a region of stable equilibrium exists at $r_{min}>0$;
    \item \label{item:unbounded} For $A \geq 0$, no local minimum exists. The potential is strictly decreasing and bounded from below. There is no bounded motion; only scattering trajectories are possible. A particle coming from infinity is reflected and returns to infinity, never remaining in a finite region.
\end{enumerate}

\begin{figure}
\resizebox{0.5\textwidth}{!}{%
  \includegraphics{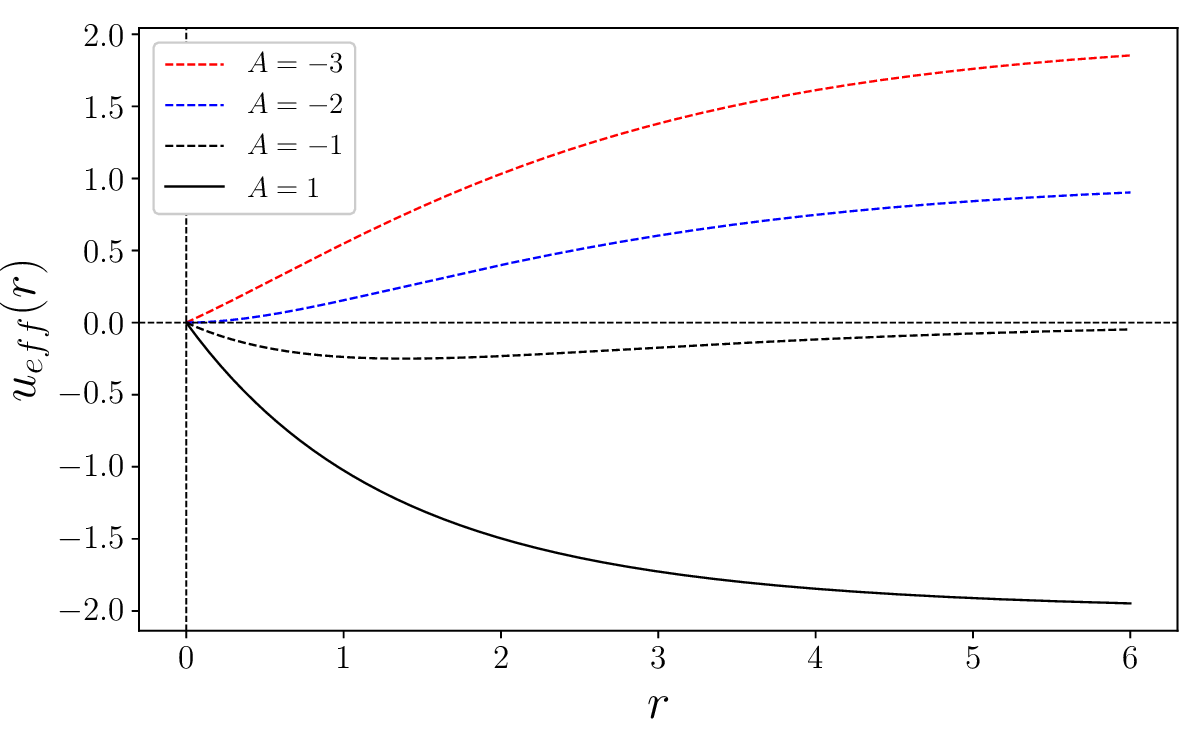}
}
\caption{The effective Hedgehog potential for different values of $A$}
\label{fig:spinA}       
\end{figure}

Because the potential approaches a finite constant at infinity, a particle can escape to infinity only if its total energy $E > -1 - Ac$. As can be viewed in figure \eqref{fig:spin}, for energies below this threshold, the particle is confined to a finite region. However, the existence of the confinement depends on the presence of a local minimum. Thus, confinement occurs only for $A<0$. In all these cases, the particle oscillates around a stable equilibrium point: either at the origin ($A < -\frac2c$) or at a finite distance ($-\frac2c < A < 0$). For $A \geq 0$, the system supports only unbounded trajectories and confinement is not possible.

\subsection{Level-dependent confinement in the Kaluza-Klein tower}
\label{ssec: spin-kk}

As in the bosonic case, once $\theta$ is treated as a compact coordinate and the system is canonically
quantized, the mass appearing in \eqref{Qr}-\eqref{effn2} is discretized as \eqref{discmass}, and the Grassmannian constant \eqref{xik} is discretized as 
$$
    \xi_{k(l)} = l \psi^\theta_k.
$$
This quantization has a distinctive consequence for the spinning particle that has no counterpart in the bosonic sector: since the confinement parameter, in the equation \eqref{spinprm}, is $A=\frac i{m^2} \xi_k \psi^r_k(0)$, it becomes explicitly level dependend,
\begin{equation}\label{eq: Al}
    A_l = i\frac{\gamma}{l}\,\psi^\theta_k\psi^r_k(0), \, l\neq0.
\end{equation}
The scaling $A_l \propto 1/l$ assumes that $\psi^\theta_k\psi^r_k(0)$, a constant of motion fixed by the initial spin data, is independent of the KK level; this way, the confinement $A<0$ established in \eqref{item:stable-origin-strict}-\eqref{item:stable-rmin} is generically satisfied only for the lightest modes of the tower, while the heavier modes ($|l|$ large) have  $|A_l| \rightarrow 0$ and asymptotically recover the unconfined, purely scattering behavior of the bosonic test particle. The mass-generation mechanism discussed here therefore does not confine the full Kaluza-Klein spectrum uniformly: it acts as a mass-dependent filter, trapping the low-lying modes near the brane while allowing the heavier modes of the same tower to escape into the bulk. The zero mode $l=0$ requires separate treatment, since \eqref{eq: Al} is singular in this limit and the expansion \eqref{effn2}-\eqref{spinprm} must be revisited directly in terms of $\dot\theta \rightarrow 0$. 

Since $U_{eff}(r) = m^2 u_{eff}(r)$ with $m^2$ a positive, $r$-independent constant at fixed level $l$, the location and stability character of the critical points of $u_{eff}$ transfer directly to $U_{eff}$; the level-dependence enters solely through $A_l$ inside $u_{eff}$ itself as anticipated above. Moreover, the approach to the unconfined regime as $|l|\rightarrow\infty$ is smooth rather than abrupt: since $r_{min} = -\frac2c \ln{(-A_l c/2)}$ and $A_l\rightarrow0^-$ as $|l|\rightarrow\infty$, the equilibrium radius diverges, $r_{min}\rightarrow\infty$, so heavier modes are not sharply expelled from the trapping region but rather confined at increasingly larger distances from the brane, continuously interpolating toward the purely scattering bosonic behavior.

\begin{figure}
\resizebox{0.5\textwidth}{!}{%
  \includegraphics{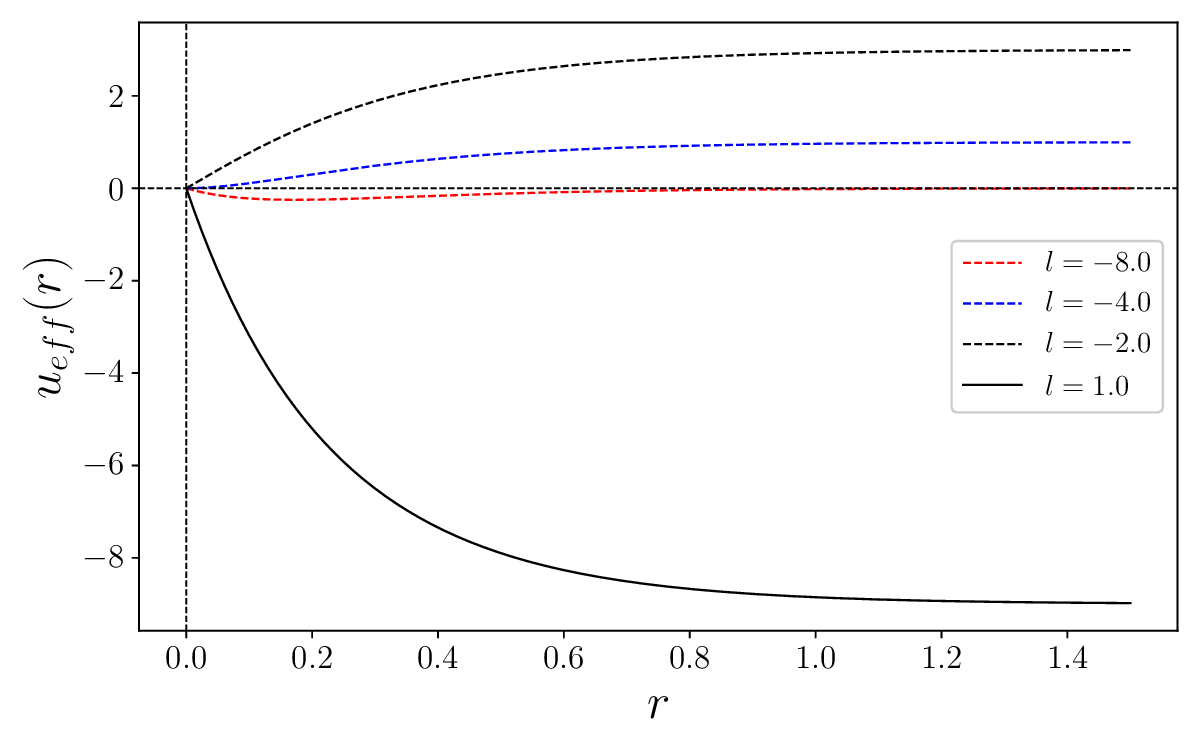}
}
\caption{The effective Hedgehog potential for different values of $l$}
\label{fig:spin}       
\end{figure}
\section{Conclusions and Perspectives}

In this work we have shown that a massless particle propagating in the bulk of a
codimension-2 hedgehog-type bran-\newline
eworld acquires an effective mass on the brane,
generated purely by its motion along the extra angular dimension,
$m=\gamma^{1/2}\dot\theta$. This mechanism was established first for a bosonic test
particle in Section \ref{confboson} and then extended to the spinning particle with $N=1,2$
pseudoclassical supersymmetry in Section \ref{confspin}, in both cases without invoking spontaneous
symmetry breaking or a scalar sector: mass here is a purely geometric consequence of
the flux-supported bulk background.

A central result of this analysis is the role played by spin. For the bosonic particle, the effective potential \eqref{ueffpform} is monotonically repulsive, admitting no
equilibrium point: the particle is never confined near the brane, only scattered. Once spin degrees of freedom are included, the coupling between spin and bulk curvature, encoded in the parameter
\begin{equation}
A = \frac{i}{m^2}\,\xi_k \psi^r_k(0),
\end{equation}
qualitatively changes this picture: a stable equilibrium point emerges for spin states with $A < 0$, where the effective potential has a minimum, either at the origin or at a finite $r_{min}>0$, while for $A\ge0$ the bosonic-like repulsive behavior is recovered.

This last observation becomes considerably sharper once $\theta$ is treated as a compact coordinate. Canonical quantization forces
\begin{equation}
\dot\theta = \frac{l}{\gamma}, \, l \in \mathbb{Z},
\end{equation}
replacing the continuous mass spectrum with a discrete Kaluza-Klein-like tower,
\begin{equation}
m_l = \frac{l}{\sqrt{\gamma}},
\end{equation}
with effective compactification radius
\begin{equation}
R_{eff} = \sqrt{\gamma} = \frac{Q}{\sqrt{2}\,c}
\end{equation}
set dynamically by the bulk flux and cosmological constant rather than imposed externally. Because the confinement parameter is inversely proportional to $m^2$, it becomes level-dependent,
\begin{equation}
A_l = \frac{i\gamma}{l}\,\psi^\theta_k \psi^r_k(0) \;\propto\; \frac{\gamma}{l},
\end{equation}
and the confinement condition $A_l<0$ is generically satisfied only for the lightest
modes. Importantly, this suppression is not a sharp cutoff: since
\begin{equation}
r_{min} = -\frac{2}{c}\ln\!\left(-\frac{A_l c}{2}\right)
\end{equation}
diverges continuously as $A_l\to0^-$, heavier modes are not abruptly expelled from the trapping region but instead settle into equilibrium at increasingly larger distances from the brane, smoothly interpolating toward the unconfined bosonic behavior as $|l|\to\infty$. The spin-curvat-\newline
ure coupling thus does not confine the
Kaluza-Klein tower uniformly: it acts as a mass-dependent filter with a continuous falloff, trapping the low-lying modes close to the brane while progressively releasing heavier modes of the same tower toward the bulk. Spin is therefore not
merely a spectator or a uniform stabilizing ingredient, as a first reading of Section \ref{confspin} alone might suggest, but the mechanism responsible for selecting,
continuously in $l$, how localized each level of the mass spectrum remains.

It is worth stressing a limitation of the present approach. The clean separation between radial and angular conserved quantities used throughout Sections \ref{confboson} and \ref{confspin} relies on the transverse metric being flat, 
\begin{equation}
dr^2 + \gamma\, d\theta^2,
\end{equation}
which holds only for codimension $n=2$. For $n>2$ the transverse space is $\gamma\, d\Omega^2_{n-1}$, an $(n-1)$-sphere with nonvanishing intrinsic curvature; the resulting Christoffel symbols couple the radial and angular sectors, so the
individual angular velocities are no longer separately conserved, and the mass-generation and mode-filtering mechanisms described here do not straightforwardly extend beyond $n=2$.

Several directions remain open. First, the zero mode $l=0$ requires dedicated
treatment, since $A_l$ is singular in this limit and the small-$r$ expansion of the effective potential must be revisited directly in terms of $\dot\theta\to0$. Second, it would be worthwhile to obtain the full localization profile analytically --
determining $r_{min}(l)$ and its rate of divergence as a function of $\gamma$ and the fermionic condensate $\psi^\theta_k\psi_k^r(0)$ -- turning the qualitative filtering picture above into a quantitative localization law for the observable brane spectrum. Third, it would be worthwhile to identify a conserved quantity that survives for $n>2$, such as the total angular momentum on the transverse sphere, and to investigate whether an
analogous -- even if not level-uniform -- filtering mechanism persists. Fourth, a full canonical quantization of the spinning sector, along the lines of the BRST treatment in \cite{Rivelles:1990dq}, would clarify whether this level-dependent confinement survives beyond the pseudoclassical approximation used here. In particular, since the
worldline fermions $\psi^A_k$ are expected to satisfy Clifford-algebra relations upon quantization, as in the standard spinning-particle construction \cite{Brink:1976uf},
it would be worth determining whether the transverse component  $\psi^\theta_k$ endows each level of the Kaluza-Klein tower with a definite spin or chirality structure
correlated with $l$, and whether this structure is what ultimately controls the level-depend-\newline
ent confinement parameter $A_l$ found in Section \ref{ssec: spin-kk}. Finally, the $N=2$ case, briefly noted in section \ref{confspin} as corresponding to a gauge field, deserves an explicit treatment connecting it to the antisymmetric tensor formulation
of~\cite{Howe:1989vn}, and it would be interesting to explore the phenomenological viability of the resulting filtered mass spectrum against existing bounds on compact and warped extra dimensions.

\section*{Statement}
The authors used an AI assistant to improve the text language and grammar and in checking computations.

\acknowledgments

I. M. Macêdo and F. E. A. de Souza are thankful for the financial support provided by the Fundação Cearense de Apoio ao Desenvolvimento Científico e Tecnológico (FUNCAP) through processes n{$^\circ$} 31052.000190/2025-40 and FPD-$0213$-$00349.01.01/23$, respectively.

%
%
%
 \bibliographystyle{unsrt}
 \bibliography{biblio.bib}
%

\end{document}